\documentclass[twocolumn,superscriptaddress, floatfix, showpacs, aps, prl, 10pt]{revtex4-2}
\usepackage{graphicx}
\usepackage{amsmath}
\usepackage[colorlinks=true, citecolor=blue, urlcolor=blue]{hyperref}
\usepackage{amssymb}
\usepackage{bm}
\usepackage{color}
\usepackage{array}
\usepackage{booktabs}
\usepackage{multirow}
\usepackage{braket}
\usepackage{xcolor}
\usepackage{soul}

\begin{document}
\title{Gapless topological Peierls-like instabilities in more than one dimension}
\author{Santiago Palumbo}
\affiliation{Instituto Balseiro, Univ. Nacional de Cuyo, Av. Bustillo, 9500, Argentina}
\author{Pablo S. Cornaglia}
\affiliation{Instituto Balseiro, Univ. Nacional de Cuyo, Av. Bustillo, 9500, Argentina}
\affiliation{Centro Atómico Bariloche, Instituto de Nanociencia y Nanotecnología (CNEA-CONICET), Av. Bustillo, 9500, Argentina}
\author{Jorge I. Facio}
\affiliation{Instituto Balseiro, Univ. Nacional de Cuyo, Av. Bustillo, 9500, Argentina}
\affiliation{Centro Atómico Bariloche, Instituto de Nanociencia y Nanotecnología (CNEA-CONICET), Av. Bustillo, 9500, Argentina}

\date{\today}

\begin{abstract}
A periodic lattice distortion that reduces the translational symmetry folds electron bands into a reduced Brillouin zone, leading to band mixing and a tendency to gap formation, as in the Peierls transition in one-dimensional systems. However, in higher dimensions, the resulting phase can present topological obstructions preventing a complete gap opening. We discuss two different mechanisms for such obstructions, emergent Weyl nodes and symmetry protected band crossings. Based on density-functional calculations, we show these mechanisms are at play in trigonal PtBi$_2$.
\end{abstract}

\maketitle

\textit{\textcolor{blue}{Introduction.}}
The coupling between electron and lattice degrees of freedom is often key to understanding the emergence of electronic and structural order in complex materials. A textbook example is the Peierls transition, where a one-dimensional (1D) lattice becomes unstable toward dimerization, resulting in a reduced translational symmetry~\cite{peierls}. 
This transition leads to the opening of a gap in the electron spectrum which lowers the system’s total energy, and stabilizes the distorted, insulating phase.
Soon after periodic lattice distortions were observed in quasi-1D systems\,\cite{PhysRevB.8.571}, they were also discovered in effectively two-dimensional (2D) materials~\cite{PhysRevLett.32.882,Wilson01031975}. 
Since then, understanding the ingredients introduced by the higher dimensionality has remained a topical question~\cite{whangbo1991hidden,PhysRevB.77.165135, PhysRevB.88.035108,rossnagel2011origin,Pouget_2024}. 

The fundamental connection between symmetries and topological phases in electronic systems~\cite{kitaev2009periodic,ryu2010topological,bradlyn2017topological,po2017symmetry,PhysRevX.7.041069} naturally raises the question of how electronic topology is modified when translational symmetry is reduced. This issue has been considered in various settings, including how charge density waves affect gapless states on the surfaces of weak topological insulators~\cite{LIU2012906}, influence the bulk electronic structure of Weyl semimetals~\cite{PhysRevB.87.161107,PhysRevLett.109.196403,PhysRevB.92.125141,gooth2019axionic,PhysRevResearch.2.042010,PhysRevB.104.174406,PhysRevB.91.121417}, and lead to the possible emergence of topological gaps~\cite{Hsu_2021,Huang_2021,doi:10.1021/acsnano.4c13478}.

The reduction of translational symmetry leads to band folding into a smaller Brillouin zone, resulting in band crossings that are generally susceptible to level repulsion and gap opening. {While this can lead to a fully gapped insulating phase, the distorted system often remains metallic due to weak coupling between the lattice distortion and certain regions of the Fermi surface~\cite{PhysRevB.72.121103,PhysRevB.77.235104,PhysRevB.80.241108,PhysRevLett.100.196402,silva2016electronic}.
A natural question is whether the tendency toward gap opening can, in contrast, encounter obstructions rooted in topological properties rather than in weak coupling.
Residual crystal symmetries can enforce protected degeneracies at specific wave vectors~\cite{PhysRevLett.115.126803,PhysRevB.102.035125,PhysRevB.102.224503,li2021charge, https://doi.org/10.1002/adma.202101591}.
Moreover, band-touching points may arise naturally in three-dimensional systems even in the absence of explicit symmetry protection~\cite{Berry1985,Murakami_2007,vanderbilt2018berry}.
Weyl and Dirac points are prominent examples of such crossings, where topological invariants protect the degeneracies and give rise to robust surface states or exotic transport phenomena~\cite{RevModPhys.90.015001}.
These features can obstruct the formation of a full gap at the Fermi level, resulting in a topologically obstructed Peierls phase, in which levels are repelled from the Fermi energy, leaving behind a pseudogap in the spectrum.

\begin{figure}
    \centering
    \includegraphics[width=\columnwidth]{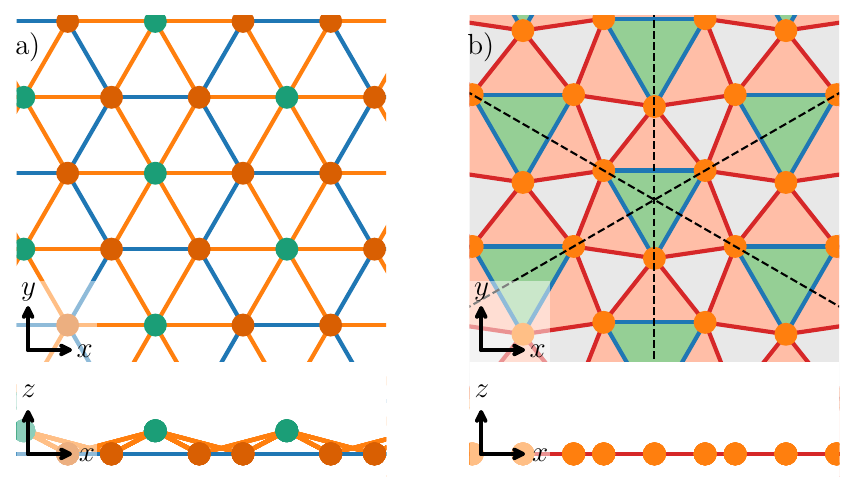}
    \caption{Examples of distortions of the triangular lattice that reduce translational symmetry while preserving threefold rotational and reflection symmetries. (a) Atoms shown as darker shaded disks are displaced out of the plane, forming a buckled, decorated honeycomb lattice. 
    (b) The equilateral triangles are distorted into triangles of varying sizes, creating sequences of alternating small and large side-sharing triangles along the mirror planes.}
    \label{fig:dim}
\end{figure}

In this Letter, we address the effects of translational symmetry reduction, focusing on systems with triangular lattice symmetry.
We demonstrate distinct mechanisms by which the electronic reconstruction associated with this symmetry reduction can stabilize a topological semimetallic phase.
Starting in two dimensions, we show that periodic lattice distortions that preserve point group symmetries can give rise to Dirac cones carrying a Berry phase, unifying within a single framework the physics of honeycomb- and Kagome-based systems.
In three dimensions, we demonstrate that inversion-breaking periodic lattice distortions can stabilize a Weyl phase rather than an insulating one.
As an example, we identify trigonal PtBi$_2$, a noncentrosymmetric Weyl semimetal~\cite{Veyrat2023,Kuibarov2023,hoffmann2024fermi,PhysRevB.110.054504,oleary2025topographyfermiarcstptbi2}.
Based on density-functional theory calculations, we show that its normal phase reflects a Peierls-like mechanism obstructed by the emergence of Weyl nodes~\footnote{We use the terminology of \textit{Peierls-like} to emphasize that the main ingredient of the problems studied is the asymmetry in electron hopping amplitudes enabled by the breaking of translational symmetry, analogous to the 1D Peierls model.}.

\textit{\textcolor{blue}{ 2D Peierls instability.}}
Our starting point is a 2D triangular lattice. In Fig. \ref{fig:dim}, we present two examples of structural distortions that reduce its translational symmetry while preserving threefold rotational and reflection symmetries present in trigonal PtBi$_2$. In these examples, sequences of short and long bonds can be identified, as in the 1D Peierls case, but the patterns can not be reduced to a mere stack of weakly coupled chains of dimerized bonds, underscoring their inherently 2D nature.  
In Fig. \ref{fig:dim}a), one out of three atoms is displaced out of the plane, resulting in a buckled, decorated honeycomb lattice. 
In Fig. \ref{fig:dim}b), one out of three atoms moves along a mirror-invariant line, leading to sequences of side-sharing small and large triangles along each of such lines. 
 
In the latter case, the resulting $\sqrt{3}\times\sqrt{3}$ supercell can be described using lattice vectors $\mathbf{a}=(3/2,-\sqrt{3}/2)$ and $\mathbf{b}=(0,\sqrt{3})$ expressed in units of the triangular lattice parameter, and sites $\mathbf{s_1}=-\lambda\,\mathbf{a}/3$, $\mathbf{s_2}=\lambda(\mathbf{a}+\mathbf{b})/3$ and $\mathbf{s_3}=-\lambda\,\mathbf{b}/3$, inside the unit cell. Here, $\lambda$ is a distortion parameter that interpolates between the triangular lattice ($\lambda=1$) and the Kagome lattice ($\lambda=1.5$).
To illustrate the main effects of this distortion on the electronic structure we consider a spinless tight-binding Hamiltonian with a single $s$ orbital on each site,
\begin{equation}
H(\mathbf{k}) = -t\, F(\mathbf{k}) + \delta\, D(\mathbf{k}),
\end{equation}
where the first term on the r.h.s. describes the nearest-neighbor hoppings and the second term introduces their change due to the structural distortion [Fig. \ref{fig:monolayer}a)]. The nonzero elements of $F(\mathbf{k})$ read
\begin{align}
F_{12}(\mathbf{k}) &= e^{-i \mathbf{k} \cdot (\mathbf{s_1} - \mathbf{s_2})}\Big(1 + e^{-i \mathbf{k} \cdot \mathbf{a}} + e^{-i \mathbf{k} \cdot (\mathbf{a} + \mathbf{b})}\Big),\nonumber\\
F_{13}(\mathbf{k}) &=F_{12}(M_x\mathbf{k}), \nonumber\\
F_{23}(\mathbf{k}) &=F_{12}(C^{-1}_3\mathbf{k}),
\end{align}
where $M_x$ and $C_3$ are, respectively, the reflection operation, which acts like $M_x: \mathbf{a} \to -\mathbf{a}-\mathbf{b}$, and the three-fold rotation $C_3: \mathbf{a} \to \mathbf{b}$. The tensor $D(\mathbf{k})$ has the same symmetry properties and
\begin{equation}
D_{12}(\mathbf{k}) = e^{-i \mathbf{k} \cdot (\mathbf{s_1} - \mathbf{s_2})}\Big(1 - e^{-i \mathbf{k} \cdot \mathbf{a}} - e^{-i \mathbf{k} \cdot (\mathbf{a} + \mathbf{b})}\Big).
\end{equation}
The case $\delta>0$ describes the situation in which the hopping amplitude along shorter (longer) bonds increases (decreases).
As detailed in the Supplementary Material (SM), 
$H(\mathbf{k})$ 
preserves time-reversal, three-fold rotational, and mirror symmetries across planes containing the rotation axis~\cite{suplement}. 

For $\delta=0$, which corresponds to the triangular lattice, the system recovers inversion ($I$) symmetry as well as additional translations, making the use of a supercell unnecessary. In this case, the band structure can be obtained by folding the bands of the triangular lattice into the reduced Brillouin zone. Although this is no longer possible in the presence of a distortion, the band structure can nevertheless partially retain the folding pattern. This can be quantified by the so-called unfolding weight $w_{nk}$, which measures the overlap between a Bloch state with wave-vector $k$ and band number $n$ and the state with the same wave-vector on the undistorted lattice~\cite{PhysRevLett.106.027002,unfolding}.  
For $\delta=0$,  folded and non-folded bands have weights $w_{nk}=0$ and $w_{nk}=1$, respectively. 

As in 1D, band folding leads to band crossings at the BZ boundary, here, e.g., at the $K$ point [Fig.~\ref{fig:monolayer}b,c)]. However, unlike in 1D, the crossing at $K$ involves three bands, reflecting that this point is equidistant from three adjacent BZ centers~\cite{PhysRevB.97.115118}. In addition, along $\Gamma$-$K$, the folded bands are doubly degenerate due to the reflection symmetry, which pairs reciprocal lattice vectors, causing bands in the associated BZs to fold to the same energy along this path in the supercell first BZ. 

Folding induced degeneracies are lifted once the translation symmetry is reduced. In particular, the three-fold crossing at $K$ is split, although not necessarily generating a full gap opening.  
That is the case for $ \delta < 0 $, as shown in Fig. \ref{fig:monolayer}d), where both the double degeneracy of the folded bands along $\Gamma$-$K$ and their crossing at $ K $ with the non-folded bands gap out. Conversely, for $ \delta > 0 $,  a Dirac cone carrying a non-trivial Berry phase emerges at $ K $ [Fig. \ref{fig:monolayer}e)].

\begin{figure}
    \centering
    \includegraphics[width=\columnwidth]{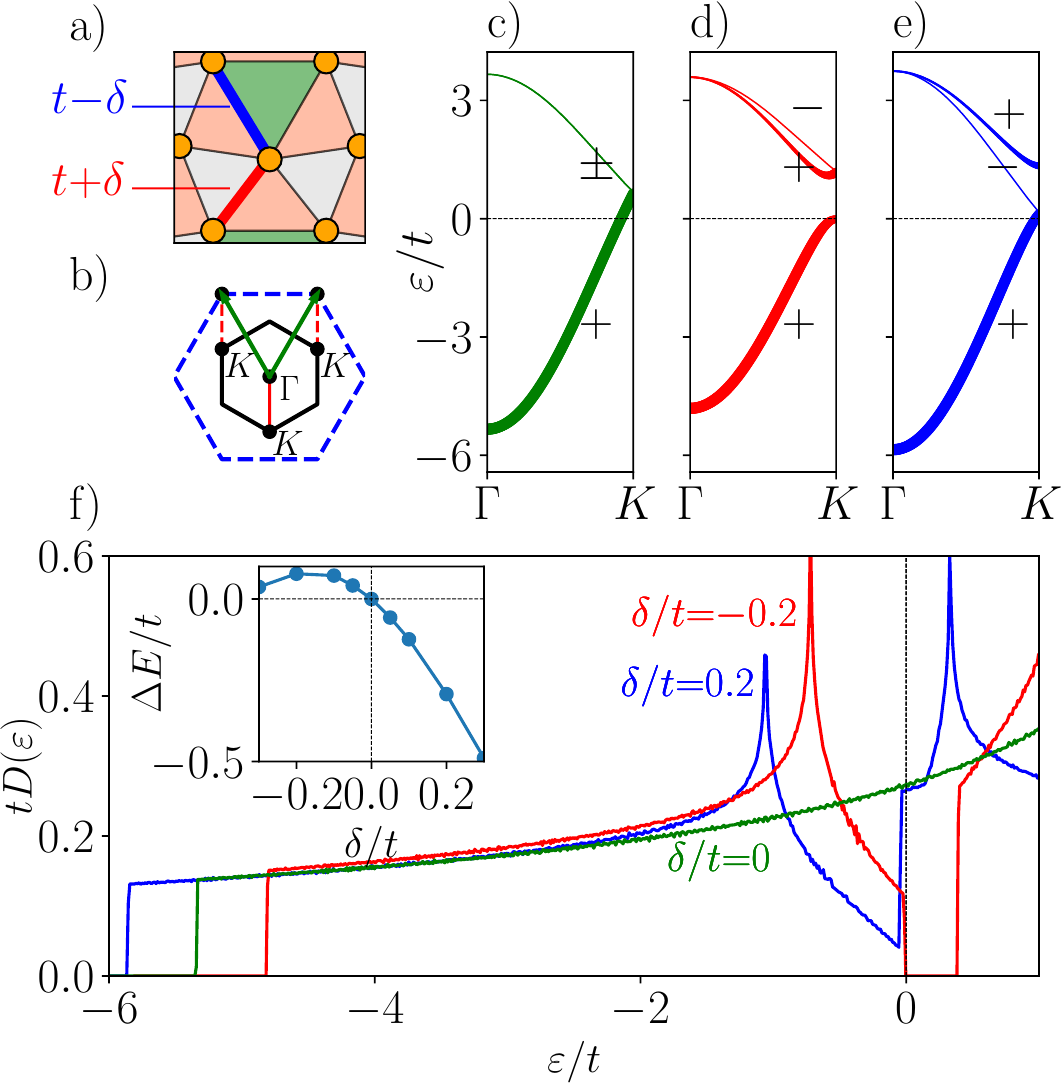}
    \caption{(a) Fragment of the distorted lattice [see Fig. \ref{fig:dim}b)] indicating the hopping amplitudes for short ($t+\delta$) and long ($t-\delta$) bonds. (b) Brillouin zones corresponding to the triangular lattice (dashed blue lines) and to the $\sqrt{3}\times\sqrt{3}$ supercell (black continous lines). The supercell Brillouin zone is rotated by $30^\circ$ and reduced in area by a factor of three relative to the original lattice. (c) Undistorted triangular lattice band structure folded into the supercell Brillouin zone. The reflection symmetry eigenvalues are indicated by $+$ and $-$ signs next to the bands. The linewidth indicates the unfolding weight which is $1$ for the lower band and $0$ for the upper bands in this case. The Fermi level ($\varepsilon_F =0$) corresponds to an occupancy of a single electron per unit cell. (d,e) Band structure for the distorted case with (d) $\delta/t = -0.2$ and (e) $\delta/t=0.2$. 
    (f) Density of states for $\delta/t =0.2$, $0$, and $0.2$. \textit{Inset:} Total electronic energy relative to the undistorted limit as a function of $\delta/t$.} \label{fig:monolayer}   
\end{figure}

These two scenarios can be classified by the mirror-symmetry eigenvalues of the Bloch states. The little group of both $\Gamma$ and $ K $ contains $ C_3 $ and reflection symmetries, allowing the formation of two-dimensional irreducible representations with states of opposite mirror eigenvalues. The two folded bands at $\Gamma$ form such a representation and have two possible ways of connecting with the states at $K$: they split in energy and pair again at $K$, or they swap, with one folded band crossing at $K$ with the non-folded band. 

Thus, the sign of $\delta$ controls a phase transition between an insulator and a symmetry-enforced topological semimetal.
Notably, the latter phase can be smoothly connected with Kagome-like systems, with a flat band and a Dirac cone at $K$ obtained in the present model in the $\delta\to t$ limit.
Similar obstructions to a full gap opening can be obtained for the distortion in Fig. \ref{fig:dim}a). 
An important difference in this case is that the point symmetry of the atomic sites includes $C_3$. 
For this reason, the distorted electronic structure is smoothly connected to the honeycomb lattice case.
These findings place in perspective a common origin of Dirac cones in Kagome and honeycomb systems: a reduction of translational symmetry relative to a triangular lattice, while preserving a non-Abelian little group at the K points.
In the SM, we present different distortions of a triangular lattice where a preserved point symmetry connecting sites of the enlarged unit cell stabilizes folding-induced band crossings~\cite{suplement}.

Fig. \ref{fig:monolayer}f) illustrates that a finite $\delta$ has a significant effect on the electronic structure near the Fermi level and on the total bandwidth.  
Since the short bonds form continuous pathways across the system,  when $\delta > 0$ their increased hopping amplitude leads to an increased  bandwidth. Conversely, as the long bonds form large triangles connected to each other only through short bonds [see Fig. \ref{fig:dim}b)], $\delta<0$ favors charge localization at these large triangles and thus a bandwidth reduction.
 
 These changes modify the total electronic energy relative to the undistorted case ($\Delta E$), as depicted in the inset of Fig. \ref{fig:monolayer}f). For both signs of $\delta$,  the strong low-energy reconstruction, from a metal at $\delta=0$ to either an insulator for $\delta<0$ or to a semimetal for $\delta>0$, transfers spectral weight from the Fermi level to lower energies, reducing the total electronic energy. The bandwidth change leads to an additional reduction of the energy for $\delta>0$ while it competes with the Fermi level effect for $\delta<0$. For small $\delta$ this results in $\Delta E \propto \delta$, showing that the electronic contributions can drive a finite distortion.

\begin{figure}
    \centering
    \includegraphics[width=\columnwidth]{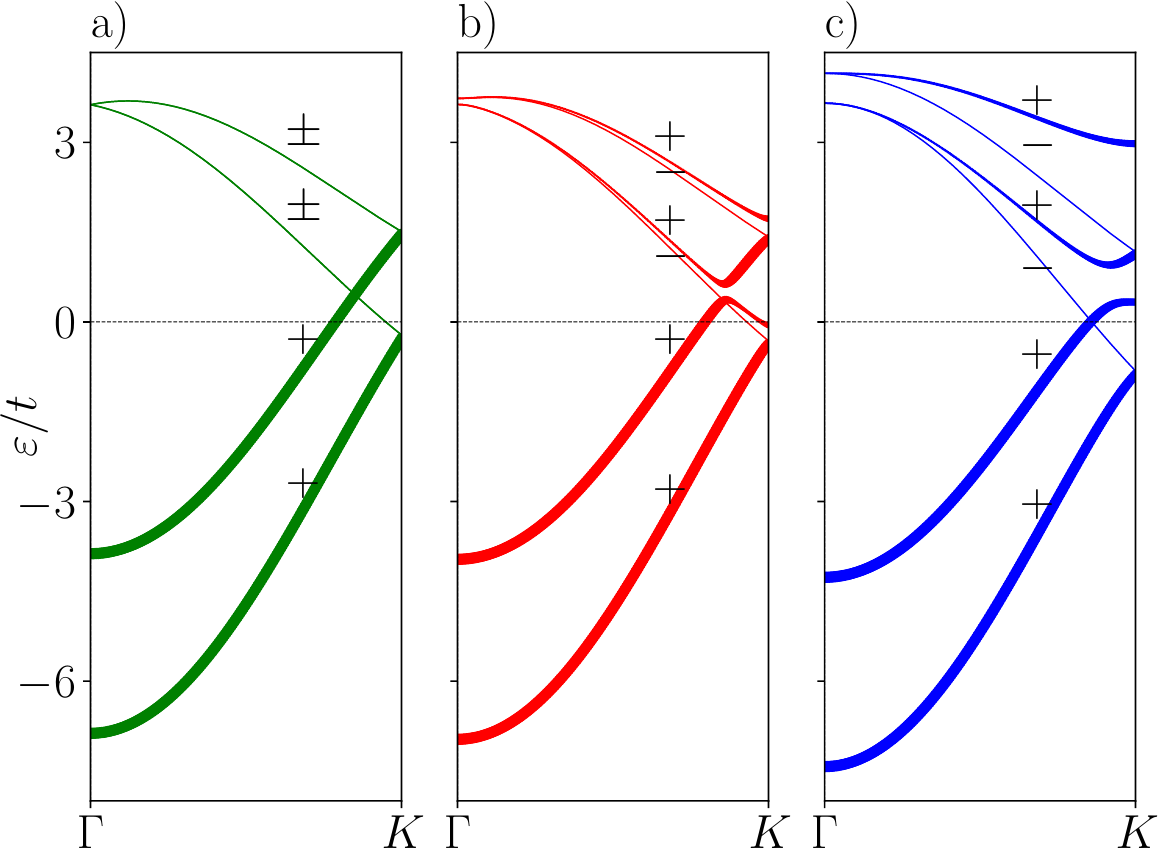}
    \caption{Band structure of a triangular AB stacked bilayer (a) without distortion and (b,c) with one of the layers dimerized: (b) $\delta/t=0.1$ and (c) $\delta/t=0.5$. Energies are measured with respect to the 1/3 filling chemical potential. As in Fig. \ref{fig:monolayer}, the linewidth indicates the unfolding weight  and the reflection symmetry eigenvalues are indicated with $+$ and $-$ signs.}
    \label{fig:bilayer}
\end{figure}

 \textit{\textcolor{blue}{The case of a bilayer.} }
Let us consider a stack of two triangular lattices, based as before on a $\sqrt{3}\times\sqrt{3}$ supercell setting.  In the absence of a distortion, the layers are connected by inversion symmetry. Atoms in one layer are positioned as in the previous section (shifted by a finite amount along the stacking direction), while those in the second layer obey 
 $\mathbf{s_{i+3}}=-\mathbf{s_i}$, with $i=\{1,2,3\}$. 
We consider the Hamiltonian 
\begin{equation}
H_{b}(\mathbf{k}) = 
\begin{pmatrix}
-t \,F(\mathbf{k}) & \nu \, C(\mathbf{k}) \\
  \nu\, C^\dagger(\mathbf{k}) & -t\, F(\mathbf{k}) 
\end{pmatrix}
 +
\begin{pmatrix}
\delta \,D(\mathbf{k}) & 0\\
  0 & 0
  \end{pmatrix}.
\end{equation}
The first term on the right hand side describes the centrosymmetric limit, including in-plane first-neighbor hoppings within each layer as well as an interlayer first-neighbor coupling described by $C(\mathbf{k})$. In order to ensure the same symmetries than in the monolayer, we have
$C_{ij}(\mathbf{k}) = g(C^{\eta_{ij}}_3 \mathbf{k})$,
where $\eta_{ij}=\epsilon_{ij} (-1)^{i+j}$, with $\epsilon_{ij}$ the 2D Levi-Civita tensor and $g(\mathbf{k})=e^{i \mathbf{k}\cdot(\mathbf{s_1}-\mathbf{s_4})}$.
The second term reduces the translation symmetry by distorting only one of the layers.

For $\delta=0$ and finite interlayer coupling $\nu$, the band structure resembles the monolayer case, yet it is effectively doubled due to formation of bonding and antibonding states. As a result, there are two three-fold crossings at $K$. Typically, the folded pair of bands associated with one of these crossings at $K$ can intersect with the non-folded band involved in the second crossing. This intersection forms a three-fold crossing that can occur at any momentum along $\Gamma$-$K$ and is also unstable under perturbations that reduce the translational symmetry. Nonetheless, as in the monolayer, any perturbation that preserves the reflection symmetry will allow the intersection between the non-folded band and its folded counterpart of opposite mirror eigenvalue to persist, as exemplified in Fig. \ref{fig:bilayer}c).

Thus,  the dimerization in a bilayer stabilizes a topological semimetal albeit in this case the topological band crossing is no longer pinned at the BZ border.  This mechanism is also relevant in three dimensional systems where the translational symmetry is reduced along two dimensions.  
For a stack of coupled bilayers, which reproduces the Bi crystal structure in PtBi$_2$, the above analysis can be easily extended to each $k_z$ plane, resulting in the formation of a mirror-protected nodal line. 

\textit{\textcolor{blue}{The case of PtBi$_2$.} }
Fig.  \ref{fig:dft}a) shows the crystal structure of PtBi$_2$ in the space group P31m (157)~\cite{ptbi2:kaiser14,Shipunov2020}. Distorted triangular lattices as those in Fig. \ref{fig:dim} are stacked along with a Pt triangular lattice every two Bi layers.
The centrosymmetric parent phase formed by undistorted triangular lattices can be stabilized by substituting Bi with Te or Sb~\cite{Takaki2022}.
The distortion linking these two structures controls a transition between a metal and a topological semimetal~\cite{us_long}.
This strong electronic reconstruction originates in the reduction of translational symmetry and, accordingly, it is weakly affected by the SOC.
We next focus on the case without SOC to demonstrate that even in its absence, the semimetallic phase is characterized by the emergence of Weyl nodes. This explicitly shows that the purely-orbital physics involved in inversion-breaking Peierls instabilities in three dimensions can stabilize a Weyl phase. A detailed discussion of the effects of SOC, along with estimates of the hopping amplitude asymmetries induced by the reduced translational symmetry, is presented in \cite{us_long}.

We perform \textit{ab-initio} calculations using the experimentally reported structure of noncentrosymmetric PtBi$_2$ \cite{Shipunov2020} ($S_0$) and of centrosymmetric Pt(Bi$_{0.901}$Te$_{0.099}$)$_2$~\cite{Takaki2022} ($S_1$), in both cases representing the systems with the space group P31m~\cite{suplement}.
This approach neglects effects of the Bi by Te substitution other than the change in the overall translation and inversion symmetries. We also consider intermediate structures obtained as $S_\alpha = (1-\alpha) S_0 + \alpha S_1,\,\,\, 0 \leq \alpha \leq 1$.

\begin{figure}
    \centering
    \includegraphics[width=\columnwidth]{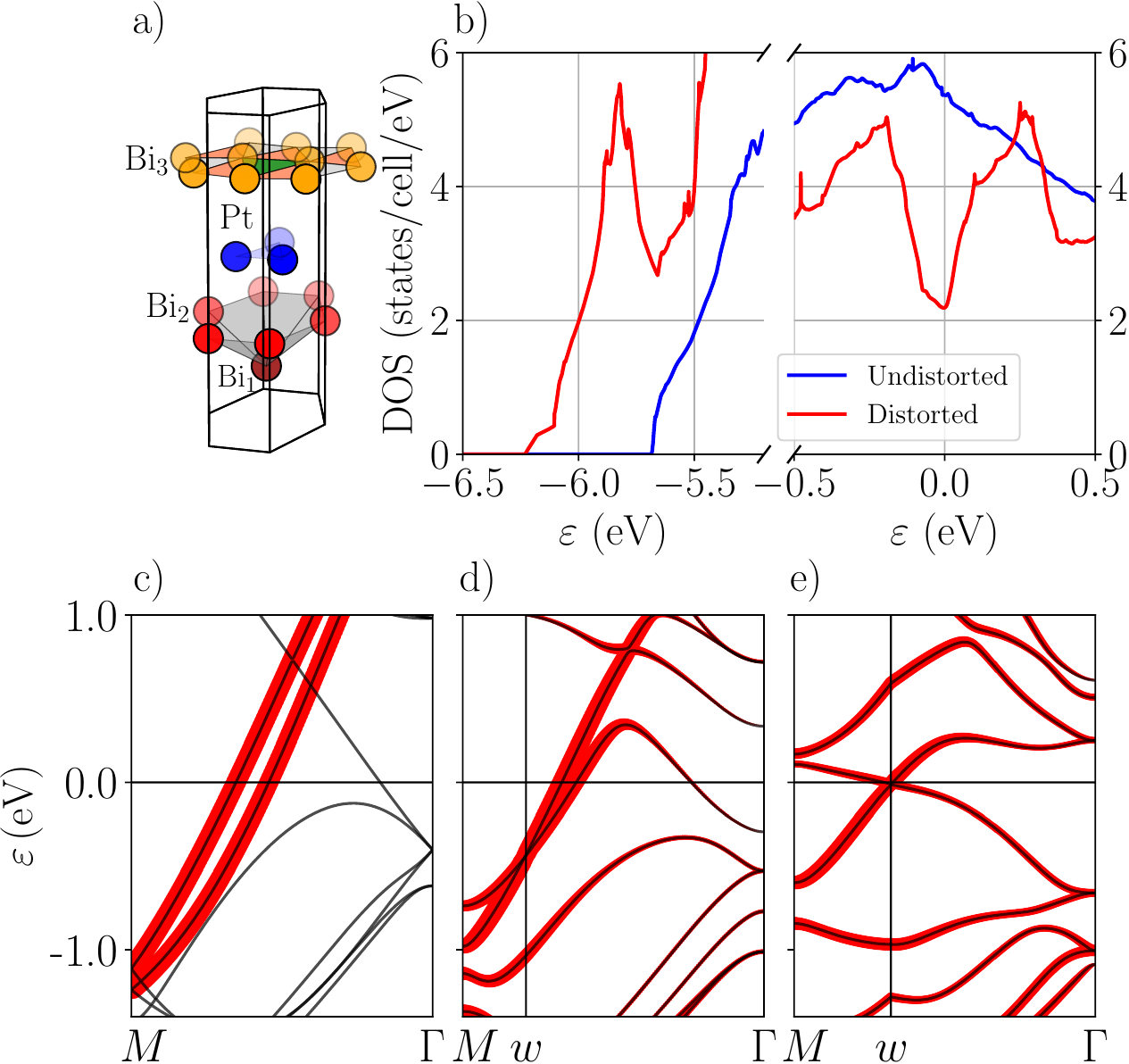}
    \caption{(a) Crystal structure of PtBi$_2$. (b) Density of states in the absence of SOC for the undistorted (space group P$\bar{3}$2/m1) and distorted (P31m) crystal structures. (c,d,e) Band structure of in the absence of spin-orbit coupling for different crystal structures, the size of the red dots indicates the unfolding weight: c) centrosymmetric limit ($\alpha=0$) where the unfolding weight exctly distinguished folded bands; d) intermediate value of $\alpha=0.39$ where Weyl nodes emerge; e) noncentrosymmetric case ($\alpha=1$).}
    \label{fig:dft} 
\end{figure}

Fig. \ref{fig:dft}b) shows the density of states (DOS), which reveals a pronounced reduction at the Fermi level in the noncentrosymmetric case compared to the centrosymmetric one, which is indicative of a metal-to-semimetal transition. In addition, the distorted structure leads to an approximately 0.5\,eV increase in the overall bandwidth. 

	Thus, as in the elementary models discussed earlier, the periodic lattice distortion in PtBi$_2$ gives rise to two distinct electronic effects: a low-energy effect, consisting of a strong depletion of spectral weight near the Fermi energy, and a high-energy effect associated with an overall enhancement of hopping amplitudes. Both contribute to lowering the total electronic energy, which we find to be 50\,meV lower in the distorted structure~\footnote{For this comparison, we consider fully relaxed crystal structures.}.

The band structure near the $\Gamma$ point illustrates how the suppresion of spectral weight takes place (Fig.~\ref{fig:dft}c-e). In the undistorted case, states from the BZ border are folded to $\Gamma$ at -0.4\,eV. The distortion splits their degeneracy, opening a gap of 0.8\,eV at this wave-vector. 
The resulting phase is yet not insulating due to the emergence of topological crossings. 
These include nodal lines within mirror-invariant planes, the origin of which follows the ideas discussed in the previous section. 
Secondly, and more interesting as it best illustrates key aspects linked to the dimensionality, the distortion also leads to the pairwise creation of Weyl nodes.

These nodes involve bands that at $\delta=0$ can be identified as folded and non-folded. They are not present for arbitrarily small distortions, rather, they emerge from the $\Gamma$-$M$ lines at a finite value of $\alpha\approx 0.39$ and persist for $\alpha$ up to $1$.  
They are monopoles of the Berry curvature which, without SOC, has as only sources orbital and site degrees for freedom~\cite{lesne2023designing,mercaldo2023orbital}.

	The case of PtBi$_2$ shows how a Weyl phase can be stabilized by a periodic lattice distortion that breaks inversion symmetry, without relying on SOC. 
While conventional level repulsion is present between folded and non-folded bands, the emergence of stable crossing points at generic crystalline momenta becomes possible in three dimensions. This behaviour is rooted in a codimensionality argument: under broken inversion symmetry the crossing of two non-degenerate bands requires satisfying three constraints~\cite{Murakami_2007}. An external control parameter, in our case the amplitude of a periodic lattice distortion, enables tuning into such a gapless phase.  Once the Weyl nodes are established, their stability is reflected in the fact that further changes to the control parameter shift their position in momentum space without opening a gap.

	This stands in contrast to systems where charge density waves lead to insulating behavior~\cite{chen2015charge,PhysRevB.95.245136}, or where metallicity is preserved due to weak coupling between low-energy states and the distortion potential~\cite{silva2016electronic}. Instead, PtBi$_2$ falls into a strong-coupling regime, where significant spectral weight is removed from the Fermi level, yet a full gap does not open due to topological constraints. This strong level repulsion, together with charge conservation, helps explain why the low-energy electronic structure is relatively uncluttered compared to the centrosymmetric limit, with the Weyl nodes lying so close to the Fermi energy (Fig.\ref{fig:dft}d). Importantly, these crossings remain robust upon inclusion of SOC\cite{us_long}, which acts to couple the nodes from each spin sector and split them in energy. Finally, since this mechanism does not rely on fine details such as lattice geometry or orbital character, it may be broadly relevant to other systems exhibiting inversion-breaking periodic lattice distortions in three dimensions.

\textit{\textcolor{blue}{Conclusions.} }
We have analyzed Peierls-like instabilities in systems with triangular lattice symmetry, where translation symmetry is reduced. As in the one-dimensional Peierls transition, the electronic reconstruction contributes to stabilizing the crystal distortion. In higher dimensions, however, the fate of the electronic structure can be constrained by a topological obstruction that prevents the opening of a full gap.
We have examined cases  where this obstruction arises from a band crossing that carries a Berry phase of $\pi$, as well as a case involving a Chern number of 1. 
Our findings establish a smooth connection with the limiting cases of both Kagome and honeycomb lattices and suggest that materials based on distorted triangular lattices may host the electronic instability we describe. As an illustrative example, we have argued that the topology of trigonal PtBi$_2$ can be effectively understood within this framework.

\textit{\textcolor{blue}{Acknowledgments. }} JF acknowledges useful discussions with Riccardo Vocaturo, Klaus Koepernik, Oleg Janson, Ion Cosma Fulga, and Jeroen van den Brink. Computational resources were provided by the HPC cluster of the Physics Department at Centro Atómico Bariloche (CNEA).

\bibliography{PtBi2.bib}

\end{document}